# E-Bayesian Estimation For Some characteristics Of Weibull Generalized Exponential Progressive Type-II Censored Samples

Hassan Piriaei[a,*] and Omid Shojaee[b]

[a] Department of Mathematics, Borujerd Branch, Islamic Azad University, Borujerd, Iran

ORCID: http://orcid.org/0000-0001-8643-4464

[b] Department of Statistics, Faculty of Mathematics and Statistics, University of Isfahan, Isfahan 81746-73441, Iran

ORCID: https://orcid.org/0000-0002-5944-4285

## ABSTRACT

Estimation of reliability and hazard rate is one of the most important problems raised in many applications especially in engineering studies as well as human lifetime. In this regard, different methods of estimation have been used. Each method exploits various tools and suffers from problems such as complexity of computations, low precision, and so forth. This study is employed the E-Bayesian method, for estimating the parameter and survival functions of the Weibull Generalized Exponential distribution. The estimators are obtained under squared error and LINEX loss functions under progressive type-II censored samples. E-Bayesian estimations are derived based on three priors of hyperparameters to investigate the influence of different priors on estimations. The asymptotic behaviours of E-Bayesian estimations have been investigated as well as relationships among them. Finally, a comparison among the maximum likelihood, Bayes, and E-Bayesian estimations are made, using real data and Monte Carlo simulation. Results show that the new method is more efficient than previous methods.

1. ## INTRODUCTION

One of the most important topics in statistical inference is the problem of estimating the parameters (or characteristics) of a distribution. In reliability theory, the survival and failure rate functions of lifetime distributions are usually estimated based on available failure times from the specifically designed engineering tests. On the other hand, some additional information, such as external conditions of operation, prior beliefs of the proficients on the parameters values and viewing of internal parameters, etc., are usually existed that are very useful for more precise estimation. Bayesian methods are often beneficial tools for modeling such additional information as a prior distribution. In practice, the parameters of the prior distribution are depended on the hyperparameters. The E-Bayesian (Expected Bayesian 'EB') estimation is a good approach to overcome this problem.

---

* Corresponding author.
Email addresses: h.piriaei@iaub.ac.ir (H. Piriaei), O.shojaee@sci.ui.ac.ir (O.Shojaee)

This topic has been of interest to many authors and has been addressed for some important distributions from various aspects. For instance, in [1], the authors have estimated the reliability of inverse generalized Weibull distribution under a new loss function by different methods and compared them. In [2], a new proportional hazard model (PHM) with its application in aeronautics is proposed and the hazard rate of the flexible Weibull distribution is adjusted. Also, [3] have studied confidence intervals for the reliability of a system based on Weibull distribution and ranked set sampling data.

Many scientists have used a combination of some useful distributions to improve the efficiency of their experiments. For example, [4] have introduced the Marshall-Olkin additive Weibull distribution and have generalized at least eleven lifetime models developing the shape of the hazard rate, including increasing, decreasing, bathtub and unimodal shapes. [5] have developed a new Weibull family of distribution that includes the Weibull-Exponential distribution as a special case which is used in monitoring industrial process. Furthermore, [6] have introduced the Lindley-Poisson distribution and have investigated its properties and applications in lifetime analysis.

Statistical methods dealing with censored data and simulation have a long history in the field of engineering experiments and life testing. For example, in [7], the authors have introduced a class of distributions which generalizes the power hazard rate distribution, by combining the linear and power hazard rate functions. They have estimated the parameters of this class of distributions, which is called the power-linear hazard rate distribution, under progressively type-II censoring. The authors in [8] have estimated the parameters of Kumaraswamy-exponential distribution based on adaptive type-II progressive censored samples. [9] have studied the Bayes analysis of some important lifetime models using Markov chain Monte Carlo (MCMC) when the data are left truncated and right-censored. Estimation of stress–strength reliability for generalized Maxwell failure distribution under progressive first failure censoring has been obtained by [10].

Recently, the E-Bayesian estimation method has been considered by many researchers. For example, the authors in [11], by considering the Exponential distribution as baseline lifetime, have derived E-Bayesian estimation of simple step–stress model based on type-II censored samples. The E-Bayesian estimation of the parameters of inverse Weibull distribution under a unified hybrid censored data has been obtained in [12]. [13] have derived the E-Bayesian estimations of the cumulative hazard rate and mean residual life of generalized inverted Exponential distribution under type-II censoring. Also, [14] have obtained the E-Bayesian estimations for the reliability and hazard rate functions of Exponential distribution under record data.

One of the noticeable newest distributions is the Weibull Generalized Exponential distribution (WGED). The WGED has been introduced by [15] and has a bathtub-shaped hazard rate which is an appropriate meaningful model for some electrical data as well as human lifetime. In [16], the authors have obtained the Bayes estimators of the parameters of WGED under censored samples. The WGED has a better level of fitting than some older distributions. Motivated by this, in this paper, we consider the WGED and estimate some characteristics of it. Since it is shown that the E-Bayesian method is more efficient than previous methods; and since the progressive type-II censoring is a more cost-effective scheme, to estimate the parameter (or characteristics) of interest of WGED, we use the E-Bayesian estimation method under progressive type-II censored samples.

This paper is mainly focused on the E-Bayesian method to estimate the parameter and hazard rate functions of WGED and to compute the reliability of series and parallel systems where the components of the system have a WGED. These estimates are obtained under asymmetric linear exponential (LINEX) and symmetric squared error loss functions (SELF). Also, we show that the asymptotic behaviours of E-Bayesian estimations under different priors of the hyperparameters are the same. In addition, it is shown that the E-Bayesian estimations have smaller mean squared errors (MSE) than the previous estimations.

Some of our basic motivations for this study are as follows:

1. The WGED has a bathtub-shaped hazard rate, and it is a proper conceptual model for some electrical data as well as human lifetime.
2. It is shown that, in some cases, the WGED provides a better fit than the Exponential, Weibull, generalized Exponential, and inverted Exponential distributions.
3. It has been proved that the E-Bayesian estimation is more efficient than previous estimations such as Bayesian and maximum likelihood (ML) methods at the same time as it is simple.

The rest of the paper is organized as follows. Section 2 is devoted to introducing the WGED and its characteristics. In sections 3 and 4, the ML, Bayes, and E-Bayes estimators for the parameter and characteristics of interest, respectively, are presented. Section 5 is discussed the properties of E-Bayesian estimation. Computations and comparisons among the E-Bayesian method and corresponding ML and Bayes estimators via real data and Monte Carlo simulation are studied in section 6. Finally, section 7 concludes the paper.

2. **WEIBULL GENERALIZED EXPONENTIAL DISTRIBUTION**

Consider the three-parameter WGED with probability density function (p.d.f.):

$$f(x; \alpha, \lambda, \theta) = \alpha\lambda\theta \exp(\lambda x)(\exp(\lambda x) - 1)^{\theta-1} \exp(-\alpha(\exp(\lambda x) - 1)^\theta), \quad x > 0, \quad \alpha, \lambda, \theta > 0. \quad (2.1)$$

Throughout this paper, it is assumed that $\lambda$ and $\theta$ are known and $\alpha$ is unknown.
The cumulative distribution function (c.d.f) is:

$$F(x; \alpha, \lambda, \theta) = 1 - \exp(-\alpha(\exp(\lambda x) - 1)^\theta), \quad x > 0, \quad \alpha, \lambda, \theta > 0, \quad (2.2)$$

and the reliability function is:

$$R(t) = 1 - F(t) = \exp(-\alpha(\exp(\lambda t) - 1)^\theta), \quad t > 0, \quad \alpha, \lambda, \theta > 0. \quad (2.3)$$

The reliability of a series and parallel system with $k$ independent components, at time $t$, respectively, are:

$$R(t; s, k) = \exp(-k\alpha(\exp(\lambda t) - 1)^\theta), \quad t > 0, \quad \alpha, \lambda, \theta > 0, \quad (2.4)$$

and

$$R(t; p, k) = \sum_{i=1}^{k}(-1)^{i-1}\binom{k}{i}\exp(-i\alpha(\exp(\lambda t) - 1)^\theta), \quad t > 0, \quad \alpha, \lambda, \theta > 0. \quad (2.5)$$

The hazard rate is:
$$h(t; \alpha, \lambda, \theta) = f(t)/R(t) = \alpha\lambda\theta \exp(\lambda t)(\exp(\lambda t) - 1)^{\theta-1}, \quad t > 0, \quad \alpha, \lambda, \theta > 0. \quad (2.6)$$

As can be seen from the relations (2.3) and (2.6), the WGED is belonging to the PHM. The authors in [17] have estimated the reliability functions of the PHM.

3. **MAXIMUM LIKELIHOOD AND BAYESIAN ESTIMATION**

   **3.1. MAXIMUM LIKELIHOOD ESTIMATION**

In this paper, we use the progressive type-II censoring sample which is an important method to obtain data in lifetime studies [18]. For the implementation of this method, suppose that $n$ independent items are put on a test. The ordered m-failures are observed under the progressive type-II censoring plan $(R_1, \ldots, R_m)$ where each $R_i \geq 0$ and $\sum_{i=1}^{m} R_i + m = n$. If the ordered m-failures are denoted by $x_{(1)}, \ldots, x_{(m)}$, (for convenience notation, are denoted by $x_1, \ldots, x_m$), then the likelihood function based on the observed sample is as follows:

$$L(\theta|\underline{x}) = A \prod_{i=1}^{m}[f(x_i; \theta)[1 - F(x_i; \theta)]^{R_i}], \quad (3.1)$$

where $A = n(n - 1 - R_1)(n - 2 - R_1 - R_2) \ldots (n - m + 1 - R_1 - R_{m-1})$.
For WGED, it follows from (2.1), (2.2), and (3.1), that:

$$L(\lambda, \alpha, \theta|\underline{x}) = A(\lambda, \theta; \underline{x}) \alpha^m \exp\{-\alpha S_m\}, \quad (3.2)$$

where $\underline{x} = (x_1 \ldots, x_m)$ and $A(\lambda, \theta; \underline{x})$ is constant, and:

$$S_m = \sum_{i=1}^{m}(R_i + 1)(\exp(\lambda x_i) - 1)^\theta. \quad (3.3)$$

The log likelihood function of (3.2) is:

$$\ell(\lambda, \alpha, \theta | \underline{x}) = \ln\left(A(\lambda, \theta; \underline{x})\right) + m\ln\alpha - \alpha S_m, \tag{3.4}$$

and then :

$$MLE(\alpha) = m/S_m. \tag{3.5}$$

According to the invariance property of MLE, the corresponding MLEs of the reliability (series and parallel systems) and hazard rate functions are obtained by (2.4), (2.5), (2.6), and (3.5) respectively, after replacing $\alpha$ with its MLE.

### 3.2. BAYESIAN ESTIMATION

According to the following gamma conjugate prior density:

$$g'(\alpha|a,b) = \frac{b^a}{\Gamma(a)} \alpha^{a-1} \exp\{-\alpha b\}, \quad \alpha > 0, \quad a > 0, \quad b > 0, \tag{3.6}$$

where $\Gamma(a)$ is the gamma function, the posterior density by (3.2) and (3.6), is:

$$g(\alpha|\underline{x}) = \frac{(b+S_m)^{m+a}}{\Gamma(m+a)} \alpha^{m+a-1} \exp\{-(b+S_m)\alpha\}, \quad \alpha > 0. \tag{3.7}$$

#### 3.2.1. Bayesian Estimations under SELF

As we know, most of the Bayesian inference procedures have been developed under the SELF. According to SELF, the Bayes estimate of α, by (3.7), is:

$$\hat{\alpha}_{Bs}(a,b) = (m+a)/(b+S_m). \tag{3.8}$$

The Bayes estimates of the reliability of series and parallel systems and hazard rate functions by (2.4), (2.5), (2.6), and (3.7) are, respectively:

$$\hat{R}_{Bss}(t; s, k) = \left(\frac{b+S_m}{b+S_m+k(\exp(\lambda t)-1)^\theta}\right)^{m+a}, \tag{3.9}$$

$$\hat{R}_{Bps}(t; p, k) = \sum_{i=1}^{k} (-1)^{i-1} \binom{k}{i} \left(\frac{b+S_m}{b+S_m+i(\exp(\lambda t)-1)^\theta}\right)^{m+a}, \tag{3.10}$$

and

$$\hat{h}_{Bs}(t) = \lambda\theta\exp(\lambda t)(\exp(\lambda t)-1)^{\theta-1}\hat{\alpha}_{Bs}. \tag{3.11}$$

#### 3.2.2. Bayesian Estimations under LINEX Loss Function

LINEX loss function have introduced by [19] and further properties of its have been investigated by [20]. The corresponding Bayes estimates based on LINEX loss function are, respectively:

$$\hat{\alpha}_{BL}(a,b) = [-(m+a)/q]\ln((b+S_m)/(b+S_m+q)), \tag{3.12}$$

$$\hat{R}_{BsL}(t; s, k) = (-1/q)\ln \sum_{j=0}^{\infty} \frac{(-q)^j}{j!} \left(\frac{b+S_m}{b+S_m+jk(\exp(\lambda t)-1)^\theta}\right)^{m+a}, \tag{3.13}$$

$$\hat{R}_{BpL}(t; p, k) = (-1/q)\ln \frac{(b+S_m)^{m+a}}{\Gamma(m+a)} \int_0^\infty \alpha^{m+a-1} e^{-\alpha(b+S_m)} e^{-q \cdot \sum_{i=1}^{k}(-1)^{i-1}\binom{k}{i} e^{-i\alpha(\exp(\lambda t)-1)^\theta}} d\alpha, \tag{3.14}$$

$$\hat{h}_{BL}(t) = \lambda\theta\exp(\lambda t)(\exp(\lambda t) - 1)^{\theta-1}\hat{\alpha}_{BL}. \tag{3.15}$$

Some of integrals in this paper, such as (3.14), cannot be solved analytically in a simple closed form. To pass this problem, we use numerical computations by the corresponding software.

## 4. E-BAYESIAN ESTIMATION

Under the assumption that $a$ and $b$ are independent with joint density function
$$\pi(a,b) = \pi_1(a)\pi_2(b),$$
with $\hat{\alpha}_B(a,b)$ being continuous,
$$\hat{\alpha}_{EB} = \iint_D \hat{\alpha}_B(a,b)\pi(a,b)\,dadb, \tag{4.1}$$

is called the E-Bayesian estimation of $\alpha$, which is assumed finite, where $D$ is the domain of $a$ and $b$, $\hat{\alpha}_B(a,b)$ is Bayesian estimation of $\alpha$ with hyperparameters $a$ and $b$, and $\pi(a,b)$ is the joint density function of $a$ and $b$ over $D$. For more details, interested readers may refer to [21].

According to [22], $a$ and $b$ should be selected such that $g'(\alpha|a,b)$ is a decreasing function of $\alpha$. The derivative of $g'(\alpha|a,b)$ with respect to $\alpha$, is:
$$\frac{dg'(\alpha|a,b)}{d\alpha} = \frac{b^a}{\Gamma(a)}\alpha^{a-2}\exp\{-\alpha b\}[(a-1) - \alpha b].$$
Therefore $g'(\alpha|a,b)$ is a decreasing function of $\alpha$, if $0 < a < 1,\ b > 0$.

E-Bayesian estimations may be derived based on the following three distributions of the hyperparameters. In each of these priors, $a$ has the beta$(u,v)$ distribution, and the distribution of $b$ is uniform, a decreasing and an increasing function in $b$, respectively. These different functions are used to investigate the influence of different prior distributions on the results of E-Bayesian estimations.

$$\begin{cases} \pi_1(a,b) = \dfrac{1}{cB(u,v)}a^{u-1}(1-a)^{v-1}, & 0 < a < 1,\ 0 < b < c \\ \pi_2(a,b) = \dfrac{2(c-b)}{c^2 B(u,v)}a^{u-1}(1-a)^{v-1}, & 0 < a < 1,\ 0 < b < c \\ \pi_3(a,b) = \dfrac{2b}{c^2 B(u,v)}a^{u-1}(1-a)^{v-1}, & 0 < a < 1,\ 0 < b < c, \end{cases} \tag{4.2}$$

where $B(u,v)$ is the beta function.

### 4.1. E-Bayesian Estimations under SELF

#### 4.1.1. E-Bayesian Estimations of α

For $\pi_1(a,b)$, and by (3.8), (4.1) and (4.2), the E-Bayesian estimation of $\alpha$ is:

$$\hat{\alpha}_{EBs1} = \frac{1}{cB(u,v)}\int_0^1\int_0^c \left(\frac{m+a}{b+S_m}\right)a^{u-1}(1-a)^{v-1}dbda = \frac{1}{c}\left(m + \frac{u}{u+v}\right)\ln\left(\frac{c+S_m}{S_m}\right). \tag{4.3}$$

Also, the E-Bayesian estimation of $\alpha$ based on $\pi_2(a,b)$ and $\pi_3(a,b)$ are, respectively:

$$\hat{\alpha}_{EBs2} = \frac{2}{c}\left(m + \frac{u}{u+v}\right)\left[\frac{c+S_m}{c}\ln\left(\frac{c+S_m}{S_m}\right) - 1\right], \tag{4.4}$$

and

$$\hat{\alpha}_{EBs3} = \frac{2}{c}\left(m + \frac{u}{u+v}\right)\left[1 - \frac{S_m}{c}\ln\left(\frac{c+S_m}{S_m}\right)\right]. \quad (4.5)$$

#### 4.1.2. E-Bayesian Estimations for the Reliability of Series System

Under $\pi_1(a,b)$, $\pi_2(a,b)$ and $\pi_3(a,b)$, the E-Bayesian estimates of the reliability of series system are derived by (3.9), (4.1) and (4.2), respectively, as:

$$\hat{R}_{EBss1} = \frac{1}{cB(u,v)}\int_0^1\int_0^c \left(\frac{b+S_m}{b+S_m+k(\exp(\lambda t)-1)^\theta}\right)^{m+a} a^{u-1}(1-a)^{v-1}dbda, \quad (4.6)$$

$$\hat{R}_{EBss2} = \frac{2}{c^2 B(u,v)}\int_0^1\int_0^c (c-b)\left(\frac{b+S_m}{b+S_m+k(\exp(\lambda t)-1)^\theta}\right)^{m+a} a^{u-1}(1-a)^{v-1}dbda, \quad (4.7)$$

$$\hat{R}_{EBss3} = \frac{2}{c^2 B(u,v)}\int_0^1\int_0^c b\left(\frac{b+S_m}{b+S_m+k(\exp(\lambda t)-1)^\theta}\right)^{m+a} a^{u-1}(1-a)^{v-1}dbda. \quad (4.8)$$

#### 4.1.3. E-Bayesian Estimations for the Reliability of Parallel System

Similarly, the E-Bayesian estimates of the reliability of parallel system under $\pi_1(a,b)$, $\pi_2(a,b)$ and $\pi_3(a,b)$, can be derived by (3.10), (4.1) and (4.2), respectively, as follows:

$$\hat{R}_{EBps1} = \frac{1}{cB(u,v)}\int_0^1\int_0^c \sum_{i=1}^{k}(-1)^{i-1}\binom{k}{i}\left(\frac{b+S_m}{b+S_m+i(\exp(\lambda t)-1)^\theta}\right)^{m+a} a^{u-1}(1-a)^{v-1}dbda, \quad (4.9)$$

$$\hat{R}_{EBps2} = \frac{2}{c^2 B(u,v)}\int_0^1\int_0^c (c-b)\sum_{i=1}^{k}(-1)^{i-1}\binom{k}{i}\left(\frac{b+S_m}{b+S_m+i(\exp(\lambda t)-1)^\theta}\right)^{m+a} a^{u-1}(1-a)^{v-1}dbda, \quad (4.10)$$

$$\hat{R}_{EBps3} = \frac{2}{c^2 B(u,v)}\int_0^1\int_0^c b\sum_{i=1}^{k}(-1)^{i-1}\binom{k}{i}\left(\frac{b+S_m}{b+S_m+i(\exp(\lambda t)-1)^\theta}\right)^{m+a} a^{u-1}(1-a)^{v-1}dbda. \quad (4.11)$$

#### 4.1.4. E-Bayesian Estimations for the Hazard Rate

Under $\pi_1(a,b)$, $\pi_2(a,b)$ and $\pi_3(a,b)$, the E-Bayesian estimates of hazard rate by (3.11), (4.1)-(4.5), can be given, respectively, as follows:

$$\hat{h}_{EBs1}(t) = \lambda\theta\exp(\lambda t)(\exp(\lambda t)-1)^{\theta-1}\hat{\alpha}_{EBs1}, \quad (4.12)$$

$$\hat{h}_{EBs2}(t) = \lambda\theta\exp(\lambda t)(\exp(\lambda t)-1)^{\theta-1}\hat{\alpha}_{EBs2}, \quad (4.13)$$

and

$$\hat{h}_{EBs3}(t) = \lambda\theta\exp(\lambda t)(\exp(\lambda t)-1)^{\theta-1}\hat{\alpha}_{EBs3}. \quad (4.14)$$

### 4.2. E-Bayesian Estimations under LINEX Loss Function

The corresponding E-Bayesian estimations under LINEX Loss Function are, respectively:

#### 4.2.1. E-Bayesian Estimations of α

$$\hat{\alpha}_{EBL1} = \frac{-1}{cq}\left(m+\frac{u}{u+v}\right)\left[(c+S_m)\ln\left(\frac{c+S_m}{c+S_m+q}\right) - S_m\ln\left(\frac{S_m}{S_m+q}\right) + q\ln\left(\frac{S_m+q}{c+S_m+q}\right)\right], \quad (4.15)$$

$$\hat{\alpha}_{EBL2} = \frac{1}{qc^2}\left(m+\frac{u}{u+v}\right)[(S_m^2 + (2q+2c)S_m + q^2 + 2cq)\ln\left(\frac{c+S_m+q}{S_m+q}\right) - (S_m^2 + 2cS_m)\ln\left(\frac{c+S_m}{S_m}\right) - \quad (4.16)$$
$$c^2\ln\left(\frac{c+S_m}{c+S_m+q}\right) - qc],$$

$$\hat{\alpha}_{EBL3} = \frac{-1}{qc^2}\left(m + \frac{u}{u+v}\right)\{[S_m{}^2 + 2qS_m + q^2]\ln(\tfrac{c+S_m+q}{S_m+q}) - S_m{}^2 \ln\left(\tfrac{c+S_m}{S_m}\right) + c^2 \ln\left(\tfrac{c+S_m}{c+S_m+q}\right) - qc\}. \quad (4.17)$$

#### 4.2.2. E-Bayesian Estimations for the Reliability of Series System

$$\hat{R}_{EBsL1} = \frac{-1}{qcB(u,v)}\int_0^1\int_0^c \ln \sum_{j=0}^{\infty} \frac{(-q)^j}{j!}\left(\frac{b+S_m}{b+S_m+jk(\exp(\lambda t)-1)^\theta}\right)^{m+a} a^{u-1}(1-a)^{v-1} db\,da, \quad (4.18)$$

$$\hat{R}_{EBsL2} = \frac{-2}{qc^2 B(u,v)}\int_0^1\int_0^c (c-b)\ln \sum_{j=0}^{\infty} \frac{(-q)^j}{j!}\left(\frac{b+S_m}{b+S_m+jk(\exp(\lambda t)-1)^\theta}\right)^{m+a} a^{u-1}(1-a)^{v-1} db\,da, \quad (4.19)$$

$$\hat{R}_{EBsL3} = \frac{-2}{qc^2 B(u,v)}\int_0^1\int_0^c b\ln \sum_{j=0}^{\infty} \frac{(-q)^j}{j!}\left(\frac{b+S_m}{b+S_m+jk(\exp(\lambda t)-1)^\theta}\right)^{m+a} a^{u-1}(1-a)^{v-1} db\,da. \quad (4.20)$$

#### 4.2.3. E-Bayesian Estimations for the Reliability of Parallel System

$$\hat{R}_{EBpL1} = \quad (4.21)$$

$$\frac{-1}{qcB(u,v)}\int_0^1\int_0^c \ln\left[\frac{(b+S_m)^{m+a}}{\Gamma(m+a)}\int_0^\infty \alpha^{m+a-1} e^{-\alpha(b+S_m)} e^{-q\cdot\sum_{i=1}^k (-1)^{i-1}\binom{k}{i} e^{-i\alpha(\exp(\lambda t)-1)^\theta}} d\alpha\right] a^{u-1}(1-a)^{v-1} db\,da,$$

$$\hat{R}_{EBpL2} = \quad (4.22)$$

$$\frac{-2}{qc^2 B(u,v)}\int_0^1\int_0^c (c-b)\ln\left[\frac{(b+S_m)^{m+a}}{\Gamma(m+a)}\int_0^\infty \alpha^{m+a-1} e^{-\alpha(b+S_m)} e^{-q\cdot\sum_{i=1}^k (-1)^{i-1}\binom{k}{i} e^{-i\alpha(\exp(\lambda t)-1)^\theta}} d\alpha\right] a^{u-1}(1-a)^{v-1} db\,da,$$

$$\hat{R}_{EBpL3} = \quad (4.23)$$

$$\frac{-2}{qc^2 B(u,v)}\int_0^1\int_0^c b\ln\left[\frac{(b+S_m)^{m+a}}{\Gamma(m+a)}\int_0^\infty \alpha^{m+a-1} e^{-\alpha(b+S_m)} e^{-q\cdot\sum_{i=1}^k (-1)^{i-1}\binom{k}{i} e^{-i\alpha(\exp(\lambda t)-1)^\theta}} d\alpha\right] a^{u-1}(1-a)^{v-1} db\,da.$$

#### 4.2.4. E-Bayesian Estimations for the Hazard Rate

$$\hat{h}_{EBL1}(t) = \lambda\theta\exp(\lambda t)(\exp(\lambda t)-1)^{\theta-1}\hat{\alpha}_{EBL1}, \quad (4.24)$$

$$\hat{h}_{EBL2}(t) = \lambda\theta\exp(\lambda t)(\exp(\lambda t)-1)^{\theta-1}\hat{\alpha}_{EBL2}, \quad (4.25)$$

$$\hat{h}_{EBL3}(t) = \lambda\theta\exp(\lambda t)(\exp(\lambda t)-1)^{\theta-1}\hat{\alpha}_{EBL3}. \quad (4.26)$$

## 5. ASYMPTOTIC BEHAVIOURS AND RELATIONS AMONG THE E-BAYESIAN ESTIMATIONS

Now, we investigate the relationships among E-Bayesian estimations. To prevent prolongation of the paper, we perform theorems only for SELF. For LINEX loss function, the results will be obtained similarly using the corresponding relations.

### 5.1. Relations among $\hat{\alpha}_{EBsi}(i=1,2,3)$:

**Theorem 5.1.** Let $0 < c < S_m$ and $\hat{\alpha}_{EBsi}$ ($i=1,2,3$) be given by (4.3)-(4.5), then:

(1)   $\hat{\alpha}_{EBs3} < \hat{\alpha}_{EBs1} < \hat{\alpha}_{EBs2}$.
(2)   $\lim_{S_m\to\infty} \hat{\alpha}_{EBs1} = \lim_{S_m\to\infty} \hat{\alpha}_{EBs2} = \lim_{S_m\to\infty} \hat{\alpha}_{EBs3}$.

Proof: Please see Appendix A.

Part (1) of Theorem 1 shows that with different priors (4.2), the E-Bayesian estimates $\hat{\alpha}_{EBsi}$ (i=1, 2, 3) are also different and can be ordered. Part (2) shows that $\hat{\alpha}_{EBsi}$ (i=1, 2, 3) are asymptotically equal when $S_m$ is sufficiently large.

### 5.2. Relations among $\hat{R}_{EBssi}$ and $\hat{R}_{EBpsi}(i = 1, 2, 3)$

**Theorem 5. 2.** Let $0 < c < S_m$, $\hat{R}_{EBssi}$ and $\hat{R}_{EBpsi}(i = 1, 2, 3)$ are given by (4.6)-(4.11), then:

(1) $\begin{cases} \hat{R}_{EBps2} < \hat{R}_{EBps1} < \hat{R}_{EBps3}. \\ \hat{R}_{EBss2} < \hat{R}_{EBss1} < \hat{R}_{EBss3}. \end{cases}$

(2) $\begin{cases} \lim_{S_m \to \infty} \hat{R}_{EBps1} = \lim_{S_m \to \infty} \hat{R}_{EBps2} = \lim_{S_m \to \infty} \hat{R}_{EBps3}. \\ \lim_{S_m \to \infty} \hat{R}_{EBss1} = \lim_{S_m \to \infty} \hat{R}_{EBss2} = \lim_{S_m \to \infty} \hat{R}_{EBss3}. \end{cases}$

Proof: Please see Appendix B.

Part (1) of Theorem 5.2 shows that with different priors (4.2), the E-Bayesian estimates $\hat{R}_{EBssi}$ and $\hat{R}_{EBpsi}$ (i=1, 2, 3) are also different and can be ordered. Part (2) shows that $\hat{R}_{EBssi}$ and $\hat{R}_{EBpsi}$ (i=1, 2, 3) are asymptotically equal when $S_m$ is sufficiently large.

### 5.3. Relations among $\hat{h}_{EBi}(i = 1, 2, 3)$ :

**Theorem 5. 3.** Let $0 < c < S_m$, $\hat{h}_{EBsi}$ (i = 1, 2, 3) are given by (4.12)-(4.14), then:

(1) $\hat{h}_{EBs3} < \hat{h}_{EBs1} < \hat{h}_{EBs2}.$
(2) $\lim_{S_m \to \infty} \hat{h}_{EBs1} = \lim_{S_m \to \infty} \hat{h}_{EBs2} = \lim_{S_m \to \infty} \hat{h}_{EBs3}.$

Proof: Please see Appendix C.

Part (1) of Theorem 5.3 shows that with different priors (4.2), the E-Bayesian estimates $\hat{h}_{EBsi}$ (i=1, 2, 3) are also different and can be ordered. Part (2) shows that $\hat{h}_{EBsi}$ (i=1, 2, 3) asymptotically equal when $S_m$ is sufficiently large.

## 6. APPLICATIONS

In this section, two examples are presented to assess the performance of the estimates associated with WGED.

### 6.1. A real Data Analysis

[23] have discussed on the electric data, which are 19 failure times (in minutes) for an insulating fluid between two electrodes subject to a voltage of 34 KV: 0.19, 0.78, 0.96, 1.31, 2.78, 3.16, 4.15, 4.67, 4.85, 6.50, 7.35, 8.01, 8.27, 12.06, 31.75, 32.52, 33.91, 36.71, and 72.89. In [16] the authors have showed that the WGED can be fitted to this electric data set. Using results of equations (3.5), (3.8)-(4.26), different estimates for the unknown parameters are presented in Table 2. The first column is dedicated to censor scheme. For example, R=(1*4, 0*11) means (1, 1, 1, 1, 0, 0, 0, 0, 0, 0, 0, 0, 0, 0, 0).
The program for estimating is as follows:

- Based on the electric data, we obtain the MLEs, of $\alpha, R(t; s, k), R(t; p, k)$, and $h(t)$.

- For given values of the prior parameters, we generate samples *a* and *b* from the beta and uniform priors (4.2), respectively.
- Based on the electric data, for given values of *a* and *b*, we obtain the Bayes and E-Bayesian estimates of $\alpha$, $R(t;s,k), R(t;p,k)$, and $h(t)$.

The values of the parameters for real data are given in Table 1.

**Table 1.** The values of the parameters for real data.

| $\lambda$ | $\theta$ | a | b | u | v | c |
|---|---|---|---|---|---|---|
| 0.022 | 1.95 | 0.3 | 0.62 | 0.13 | 2 | 1.12 |

It is clear from Table 2 that, the E-Bayesian estimates are robust and satisfy the Theorems 5.1, 5.2 and 5.3. Also, the E-Bayesian estimates tend to be closer to the Bayesian estimates.

**Table 2.** Estimates of the parameters based on real data of WGED.

| m ,R | MLE | BS | BL | EBS1 | EBL1 | EBS2 | EBL2 | EBS3 | EBL3 |
|---|---|---|---|---|---|---|---|---|---|
| $\alpha$ (q=1) | | | | | | | | | |
| m=10; R=(4,4,1,0*7) | 0.5046544 | 0.5040238 | 0.4920799 | 0.4939043 | 0.4821611 | 0.4984300 | 0.4864750 | 0.4893785 | 0.4778472 |
| m=15; R=(1*4,0*11) | 0.7551532 | 0.7469420 | 0.7292817 | 0.7376206 | 0.7201228 | 0.7443637 | 0.7265510 | 0.7308776 | 0.7136946 |
| m=19; R=(0*19) | 0.9562884 | 0.9419926 | 0.9197259 | 0.9332954 | 0.9111610 | 0.9418252 | 0.9192925 | 0.9247657 | 0.9030294 |
| $R(t;s,k)$ (q=2, t=8, k=5) | | | | | | | | | |
| m=10; R=(4,4,1,0*7) | 0.9035096 | 0.9070496 | 0.8748089 | 0.9059035 | 0.8739527 | 0.9050869 | 0.8733484 | 0.9067198 | 0.8745567 |
| m=15; R=(1*4,0*11) | 0.8591299 | 0.8645737 | 0.8422885 | 0.8627888 | 0.8408187 | 0.8616302 | 0.8398686 | 0.8639468 | 0.8417681 |
| m=19; R=(0*19) | 0.8250787 | 0.8319167 | 0.8151976 | 0.8296637 | 0.8132397 | 0.8282548 | 0.8120185 | 0.8310727 | 0.8144606 |
| $R(t;p,k)$ (q=2, t=8, k=5) | | | | | | | | | |
| m=10; R=(4,4,1,0*7) | 0.912545 | 0.91612 | 0.883557 | 0.914963 | 0.882692 | 0.914138 | 0.882082 | 0.915787 | 0.883302 |
| m=15; R=(1*4,0*11) | 0.867721 | 0.873219 | 0.850711 | 0.871417 | 0.849227 | 0.870247 | 0.848267 | 0.872586 | 0.850186 |
| m=19; R=(0*19) | 0.833329 | 0.840236 | 0.82335 | 0.83796 | 0.821372 | 0.836537 | 0.820139 | 0.839383 | 0.822605 |
| $h(t)$ (q=1, t=100) | | | | | | | | | |
| m=10; R=(4,4,1,0*7) | 1.412937 | 1.411171 | 1.377730 | 1.382838 | 1.349960 | 1.395509 | 1.362038 | 1.370167 | 1.337882 |
| m=15; R=(1*4,0*11) | 2.114286 | 2.091296 | 2.041851 | 2.065198 | 2.016207 | 2.084077 | 2.034205 | 2.046319 | 1.998210 |
| m=19; R=(0*19) | 2.677426 | 2.637401 | 2.575058 | 2.613050 | 2.551078 | 2.636932 | 2.573845 | 2.589169 | 2.528311 |

### 6.2. Monte Carlo simulation

In this section, a Monte Carlo simulation in the following steps is used for computation and comparison of the MLE, Bayes, and E-Bayesian estimations:

- For given values of the prior parameters $(u,v)$, and $(0,c)$, we generate *a* and *b* from the beta and uniform prior densities (4.2), respectively.
- For given values of $(a,b)$ we generate $\alpha$ from the gamma prior density (3.6).
- For known values of $\lambda$, progressive type-II censored samples of different sizes are generated from the WGED with p.d.f (2.1).
- The estimates of $\alpha, R(t;s,k), R(t;p,k)$, and $h(t)$ are computed by equations given in the corresponding above sections.
- The above steps are repeated 5000 times and the MSEs of estimates are computed by:

$$MSE(\hat{Q}) = \frac{1}{5000} \sum E[(\hat{Q} - Q)]^2,$$

where $\hat{Q}$ is an estimate of $Q$.

The values of the parameters for Monte Carlo simulation are given in Table 3.

**Table 3.** The values of the parameters for Monte Carlo simulation.

| α | λ | θ | a | b | u | v | c |
|---|---|---|---|---|---|---|---|
| 0.9570615 | 3 | 2.5 | 0.4919733 | 0.5308612 | 0.13 | 2 | 1.12 |

The results are presented in Table 4-Table 7. Also, the corresponding figures are depicted in Figure 1 - Figure 4 It is clear from Table 4-Table 7 and Figure 1 - Figure 4 that the E-Bayesian estimates, are robust and satisfy the Theorems 5.1, 5.2, and 5.3.

**Table 4.** Estimates of $\alpha$ and their MSE based on WGED and Monte Carlo simulation (q=1).

| n, m ,R | MLE | BS | BL | EBS1 | EBL1 | EBS2 | EBL2 | EBS3 | EBL3 |
|---|---|---|---|---|---|---|---|---|---|
| n=20, m=10; R=(4,4,2,0*7) | 1.0624295 | 1.0487666 | 0.9951230 | 1.0038886 | 0.9525508 | 1.0245632 | 0.9709781 | 0.9832139 | 0.9341235 |
| | 0.15116386 | 0.12511340 | 0.09421697 | 0.10913560 | 0.08496838 | 0.12162578 | 0.09251680 | 0.09799324 | 0.07842493 |
| n=20, m=15; R=(2,2,1,0*12) | 1.0148799 | 1.0093206 | 0.9758401 | 0.9797886 | 0.9473193 | 0.9924985 | 0.9591510 | 0.9670787 | 0.9354876 |
| | 0.08031438 | 0.07251388 | 0.06084244 | 0.06614580 | 0.05698668 | 0.07058401 | 0.05995915 | 0.06213595 | 0.05437732 |
| n=20, m=20; R=(0*20) | 1.0044823 | 1.0011412 | 0.9763639 | 0.9789076 | 0.9547037 | 0.9882626 | 0.9635819 | 0.9695526 | 0.9458255 |
| | 0.05515184 | 0.05139057 | 0.04492122 | 0.04766107 | 0.04251779 | 0.05006708 | 0.04422210 | 0.04546904 | 0.04100403 |
| n=30, m=15; R=(5*3,0*12) | 1.0270646 | 1.0210275 | 0.9868171 | 0.9911423 | 0.9579655 | 1.0041343 | 0.9700525 | 0.9781502 | 0.9458786 |
| | 0.082403 | 0.074293 | 0.061692 | 0.067184 | 0.057190 | 0.071980 | 0.060454 | 0.062833 | 0.054303 |
| n=30, m=20; R=(4,4,2,0*17) | 1.00766 | 1.004154 | 0.979177 | 0.98185 | 0.957451 | 0.991283 | 0.966399 | 0.972416 | 0.948503 |
| | 0.058621 | 0.054432 | 0.047402 | 0.050437 | 0.044767 | 0.053068 | 0.046655 | 0.048028 | 0.043075 |
| n=30, m=30; R=(0*30) | 0.991515 | 0.989798 | 0.973555 | 0.974975 | 0.958987 | 0.98108 | 0.964888 | 0.968869 | 0.953086 |
| | 0.036393 | 0.034832 | 0.031809 | 0.033027 | 0.030557 | 0.034135 | 0.031395 | 0.032004 | 0.029798 |
| n=40, m=20; R=(5*4,0*16) | 1.000748 | 0.997456 | 0.972805 | 0.975306 | 0.951226 | 0.984614 | 0.960058 | 0.965998 | 0.942393 |
| | 0.057018 | 0.053123 | 0.046625 | 0.049467 | 0.044291 | 0.051887 | 0.046006 | 0.04726 | 0.042766 |
| n=40, m=30; R=(4,4,2,0*27) | 0.991642 | 0.989936 | 0.973699 | 0.97511 | 0.959129 | 0.981213 | 0.965028 | 0.969008 | 0.953231 |
| | 0.035647 | 0.034113 | 0.031131 | 0.032326 | 0.029896 | 0.033419 | 0.030720 | 0.031319 | 0.029151 |
| n=40, m=40; R=(0*40) | 0.981632 | 0.980603 | 0.968632 | 0.969529 | 0.957700 | 0.974020 | 0.962080 | 0.965039 | 0.953321 |
| | 0.025645 | 0.02486 | 0.023252 | 0.023886 | 0.022572 | 0.024469 | 0.023021 | 0.023347 | 0.022166 |
| n=50, m=25; R=(5*5,0*20) | 0.996649 | 0.994343 | 0.974713 | 0.976554 | 0.957292 | 0.983945 | 0.964384 | 0.969163 | 0.9502 |
| | 0.043893 | 0.041562 | 0.037292 | 0.03906 | 0.035609 | 0.040635 | 0.036769 | 0.037614 | 0.034566 |
| n=50, m=40; R=(4,4,2,0*37) | 0.983373 | 0.982304 | 0.970277 | 0.971209 | 0.959326 | 0.975721 | 0.963725 | 0.966698 | 0.954927 |
| | 0.027053 | 0.026214 | 0.024494 | 0.025171 | 0.023750 | 0.025798 | 0.024237 | 0.024590 | 0.023305 |
| n=50, m=50; R=(0*50) | 0.976042 | 0.975326 | 0.965838 | 0.966487 | 0.957089 | 0.970042 | 0.960574 | 0.962931 | 0.953604 |
| | 0.020612 | 0.020115 | 0.019085 | 0.019493 | 0.018646 | 0.019866 | 0.018937 | 0.019148 | 0.018382 |

From Table 4, the MSEs of all estimates significantly decrease as *n* increases. The MSEs of the ML, Bayesian and E-Bayesian estimates are ordered as follows:

$$\text{MSE}(\hat{\alpha}_{\text{EBs3}}) < \text{MSE}(\hat{\alpha}_{\text{EBs1}}) < \text{MSE}(\hat{\alpha}_{\text{EBs2}}) < \text{MSE}(\hat{\alpha}_{\text{Bs}}) < \text{MSE}(\hat{\alpha}_{\text{ML}}),$$

$$\text{MSE}(\hat{\alpha}_{\text{EBL3}}) < \text{MSE}(\hat{\alpha}_{\text{EBL1}}) < \text{MSE}(\hat{\alpha}_{\text{EBL2}}) < \text{MSE}(\hat{\alpha}_{\text{BL}}) < \text{MSE}(\hat{\alpha}_{\text{ML}}).$$

The E-Bayesian estimates have smaller MSEs than the Bayesian estimates. By increasing *n*, the E-Bayesian estimates have the smallest MSEs compared to their corresponding Bayesian and ML estimates and can be seen in Figure 1.

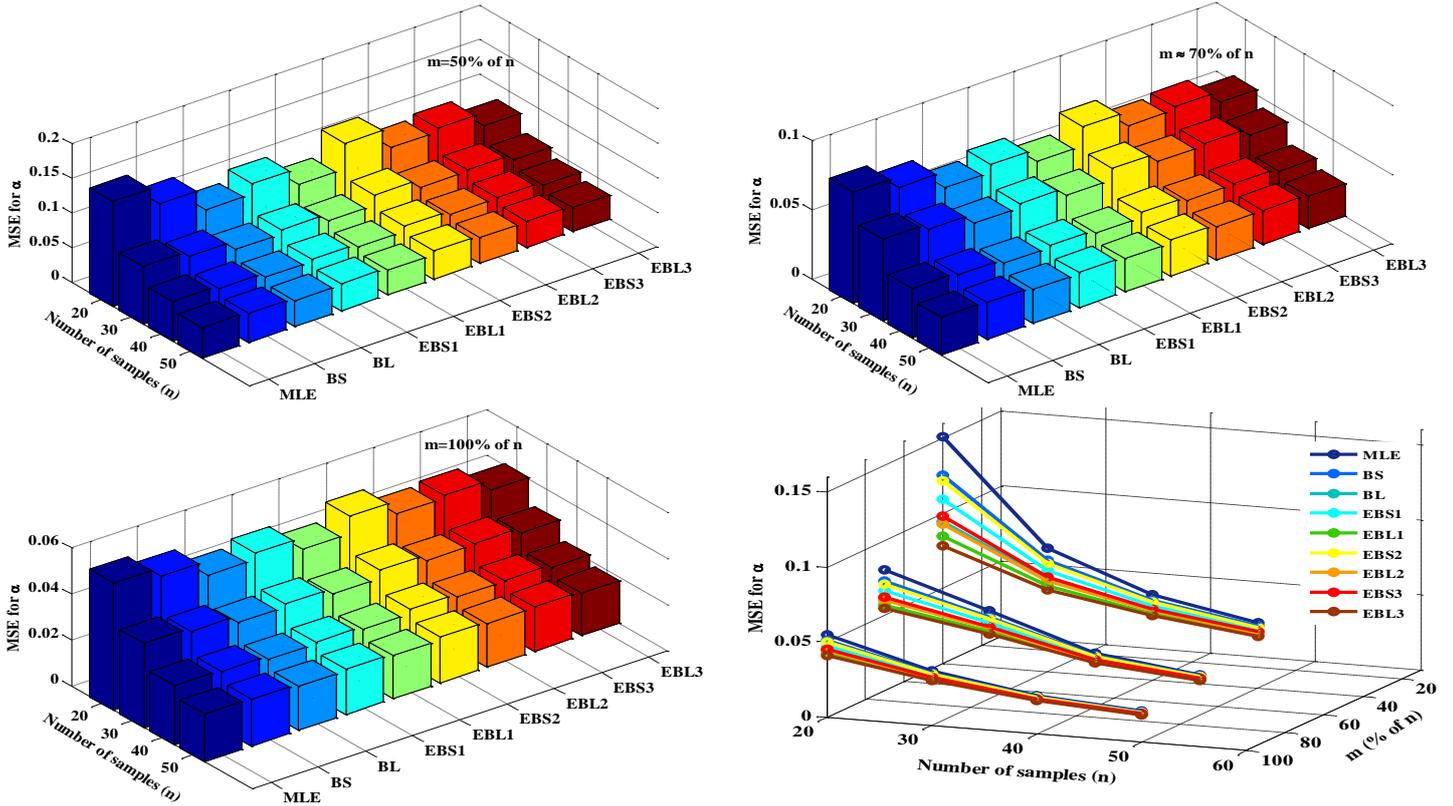

**Figure 1:** MSEs of estimations of $\alpha$ based on (50, 70 and 100)% of samples.

**Table 5.** Estimates of $R(t;s,k)$ and their MSE based on WGED and Monte Carlo simulation ($q$=2, $t$=0.1, $k$=5, $Rs$=0.7071934).

| n, m ,R | MLE | BS | BL | EBS1 | EBL1 | EBS2 | EBL2 | EBS3 | EBL3 |
|---|---|---|---|---|---|---|---|---|---|
| n=20, m=10; R=(4,4,2,0*7) | 0.686154 | 0.693364 | 0.682191 | 0.704209 | 0.692533 | 0.699523 | 0.687911 | 0.708895 | 0.697154 |
|  | 0.007356 | 0.005878 | 0.006164 | 0.005391 | 0.005428 | 0.005791 | 0.005943 | 0.005046 | 0.004968 |
| n=20, m=15; R=(2,2,1,0*12) | 0.694276 | 0.698476 | 0.689613 | 0.705785 | 0.696586 | 0.702733 | 0.693605 | 0.708836 | 0.699566 |
|  | 0.004625 | 0.004015 | 0.004074 | 0.003790 | 0.003717 | 0.003969 | 0.003951 | 0.003634 | 0.003504 |
| n=20, m=20; R=(0*20) | 0.697766 | 0.700699 | 0.693036 | 0.706215 | 0.698299 | 0.703951 | 0.696100 | 0.708478 | 0.700499 |
|  | 0.003313 | 0.002993 | 0.002987 | 0.002864 | 0.002771 | 0.002962 | 0.002904 | 0.002777 | 0.002649 |
| n=30, m=15; R=(5*3,0*12) | 0.695518 | 0.699655 | 0.69075 | 0.706940 | 0.697695 | 0.703910 | 0.694737 | 0.709969 | 0.700653 |
|  | 0.004624 | 0.004025 | 0.00406 | 0.003816 | 0.003717 | 0.003987 | 0.003945 | 0.003666 | 0.003511 |
| n=30, m=20; R=(4,4,2,0*17) | 0.698238 | 0.701145 | 0.693482 | 0.706655 | 0.698739 | 0.704399 | 0.696548 | 0.708910 | 0.700930 |
|  | 0.003238 | 0.002926 | 0.002916 | 0.002803 | 0.002708 | 0.002898 | 0.002837 | 0.002721 | 0.00259 |
| n=30, m=30; R=(0*30) | 0.701326 | 0.70314 | 0.696687 | 0.706835 | 0.700214 | 0.705344 | 0.698773 | 0.708326 | 0.701654 |
|  | 0.002196 | 0.002057 | 0.002016 | 0.001999 | 0.001910 | 0.002043 | 0.001973 | 0.001960 | 0.001852 |
| n=40, m=20; R=(5*4,0*16) | 0.696994 | 0.699966 | 0.692323 | 0.705492 | 0.697597 | 0.703216 | 0.695385 | 0.707768 | 0.699809 |
|  | 0.003419 | 0.003081 | 0.003088 | 0.002942 | 0.002863 | 0.003047 | 0.003003 | 0.002849 | 0.002735 |
| n=40, m=30; R=(4,4,2,0*27) | 0.701648 | 0.703444 | 0.697006 | 0.707137 | 0.700531 | 0.705651 | 0.699096 | 0.708623 | 0.701966 |
|  | 0.002058 | 0.001928 | 0.001891 | 0.001875 | 0.001789 | 0.001915 | 0.001848 | 0.001839 | 0.001735 |
| n=40, m=40; R=(0*40) | 0.701194 | 0.702536 | 0.696779 | 0.705329 | 0.699447 | 0.704205 | 0.698363 | 0.706453 | 0.700531 |
|  | 0.001627 | 0.001546 | 0.001526 | 0.001504 | 0.001452 | 0.001532 | 0.001493 | 0.001478 | 0.001414 |
| n=50, m=25; R=(5*5,0*20) | 0.699177 | 0.701451 | 0.694529 | 0.705885 | 0.698762 | 0.704078 | 0.697011 | 0.707692 | 0.700513 |
|  | 0.002721 | 0.002508 | 0.002487 | 0.002417 | 0.002334 | 0.002485 | 0.002426 | 0.002356 | 0.002248 |
| n=50, m=40; R=(4,4,2,0*37) | 0.70257 | 0.703883 | 0.698079 | 0.706665 | 0.700735 | 0.705551 | 0.699661 | 0.707779 | 0.701808 |
|  | 0.001595 | 0.001519 | 0.001485 | 0.001485 | 0.001419 | 0.00151 | 0.001456 | 0.001463 | 0.001384 |
| n=50, m=50; R=(0*50) | 0.703195 | 0.704226 | 0.698812 | 0.706457 | 0.700943 | 0.705567 | 0.700087 | 0.707347 | 0.701799 |
|  | 0.001258 | 0.001209 | 0.001182 | 0.001186 | 0.001135 | 0.001202 | 0.001161 | 0.001171 | 0.001111 |

**Table 6.** Estimates of $R(t;p,k)$ and their MSE based on WGED and Monte Carlo simulation ($q=2$, $t=0.25$, $k=5$, $Rp=0.8106066$).

| n, m, R | MLE | BS | BL | EBS1 | EBL1 | EBS2 | EBL2 | EBS3 | EBL3 |
|---|---|---|---|---|---|---|---|---|---|
| n=20, m=10; R=(4,4,2,0*7) | 0.786491 | 0.794755 | 0.781948 | 0.807186 | 0.793802 | 0.801815 | 0.788505 | 0.812557 | 0.799099 |
| | 0.007282 | 0.005819 | 0.006102 | 0.005337 | 0.005374 | 0.005733 | 0.005884 | 0.004996 | 0.004918 |
| n=20, m=15; R=(2,2,1,0*12) | 0.795800 | 0.800614 | 0.790455 | 0.808992 | 0.798448 | 0.805494 | 0.795031 | 0.812489 | 0.801864 |
| | 0.004579 | 0.003975 | 0.004033 | 0.003752 | 0.003680 | 0.003929 | 0.003911 | 0.003598 | 0.003469 |
| n=20, m=20; R=(0*20) | 0.799801 | 0.803163 | 0.794379 | 0.809485 | 0.800412 | 0.806890 | 0.797891 | 0.812079 | 0.802933 |
| | 0.003280 | 0.002963 | 0.002957 | 0.002835 | 0.002743 | 0.002932 | 0.002875 | 0.002749 | 0.002623 |
| n=30, m=15; R=(5*3,0*12) | 0.797224 | 0.801966 | 0.791759 | 0.810316 | 0.799719 | 0.806843 | 0.796329 | 0.813788 | 0.803110 |
| | 0.004578 | 0.003985 | 0.004019 | 0.003778 | 0.003680 | 0.003947 | 0.003906 | 0.003629 | 0.003476 |
| n=30, m=20; R=(4,4,2,0*17) | 0.800342 | 0.803674 | 0.794890 | 0.809989 | 0.800916 | 0.807404 | 0.798405 | 0.812574 | 0.803427 |
| | 0.003206 | 0.002897 | 0.002887 | 0.002775 | 0.002681 | 0.002869 | 0.002809 | 0.002694 | 0.002564 |
| n=30, m=30; R=(0*30) | 0.803881 | 0.805960 | 0.798564 | 0.810196 | 0.802607 | 0.808487 | 0.800955 | 0.811905 | 0.804257 |
| | 0.002174 | 0.002036 | 0.001996 | 0.001979 | 0.001891 | 0.002023 | 0.001953 | 0.001940 | 0.001833 |
| n=40, m=20; R=(5*4,0*16) | 0.798916 | 0.802322 | 0.793562 | 0.808656 | 0.799607 | 0.806048 | 0.797071 | 0.811265 | 0.802142 |
| | 0.003385 | 0.003050 | 0.003057 | 0.002913 | 0.002834 | 0.003017 | 0.002973 | 0.002821 | 0.002708 |
| n=40, m=30; R=(4,4,2,0*27) | 0.804250 | 0.806309 | 0.798929 | 0.810542 | 0.802970 | 0.808839 | 0.801325 | 0.812245 | 0.804615 |
| | 0.002037 | 0.001909 | 0.001872 | 0.001856 | 0.001771 | 0.001896 | 0.001830 | 0.001821 | 0.001718 |
| n=40, m=40; R=(0*40) | 0.803730 | 0.805268 | 0.798669 | 0.808470 | 0.801727 | 0.807181 | 0.800485 | 0.809758 | 0.802970 |
| | 0.001611 | 0.001531 | 0.001511 | 0.001489 | 0.001437 | 0.001517 | 0.001478 | 0.001463 | 0.001400 |
| n=50, m=25; R=(5*5,0*20) | 0.801418 | 0.804024 | 0.796090 | 0.809107 | 0.800942 | 0.807036 | 0.798935 | 0.811178 | 0.802949 |
| | 0.002694 | 0.002483 | 0.002462 | 0.002393 | 0.002311 | 0.002460 | 0.002402 | 0.002332 | 0.002226 |
| n=50, m=40; R=(4,4,2,0*37) | 0.805307 | 0.806812 | 0.800159 | 0.810001 | 0.803204 | 0.808724 | 0.801973 | 0.811278 | 0.804434 |
| | 0.001579 | 0.001504 | 0.001470 | 0.001470 | 0.001405 | 0.001495 | 0.001441 | 0.001448 | 0.001370 |
| n=50, m=50; R=(0*50) | 0.806024 | 0.807205 | 0.801000 | 0.809763 | 0.803442 | 0.808742 | 0.802461 | 0.810783 | 0.804423 |
| | 0.001245 | 0.001197 | 0.001170 | 0.001174 | 0.001124 | 0.001190 | 0.001149 | 0.001159 | 0.001100 |

From Table 5 and Table 6, the MSEs of all estimates decreases as *n* increases. The MSEs of estimates are ordered as follows:

$$\text{MSE}(\hat{R}_{EBss3}) < \text{MSE}(\hat{R}_{EBss1}) < \text{MSE}(\hat{R}_{EBss2}) < \text{MSE}(\hat{R}_{Bss}) < \text{MSE}(\hat{R}_{MLs}),$$

$$\text{MSE}(\hat{R}_{EBps3}) < \text{MSE}(\hat{R}_{EBps1}) < \text{MSE}(\hat{R}_{EBps2}) < \text{MSE}(\hat{R}_{Bps}) < \text{MSE}(\hat{R}_{MLp}),$$

$$\text{MSE}(\hat{R}_{EBsL3}) < \text{MSE}(\hat{R}_{EBsL1}) < \text{MSE}(\hat{R}_{EBsL2}) < \text{MSE}(\hat{R}_{BsL}) < \text{MSE}(\hat{R}_{MLs}),$$

$$\text{MSE}(\hat{R}_{EBpL3}) < \text{MSE}(\hat{R}_{EBpL1}) < \text{MSE}(\hat{R}_{EBpL2}) < \text{MSE}(\hat{R}_{BpL}) < \text{MSE}(\hat{R}_{MLp}).$$

The E-Bayesian estimates have smaller MSEs than the Bayesian estimates. By increasing *n*, the E-Bayesian estimates have the smallest MSEs compared to their corresponding Bayesian and ML estimates and can be seen in Figure 2 and Figure 3.

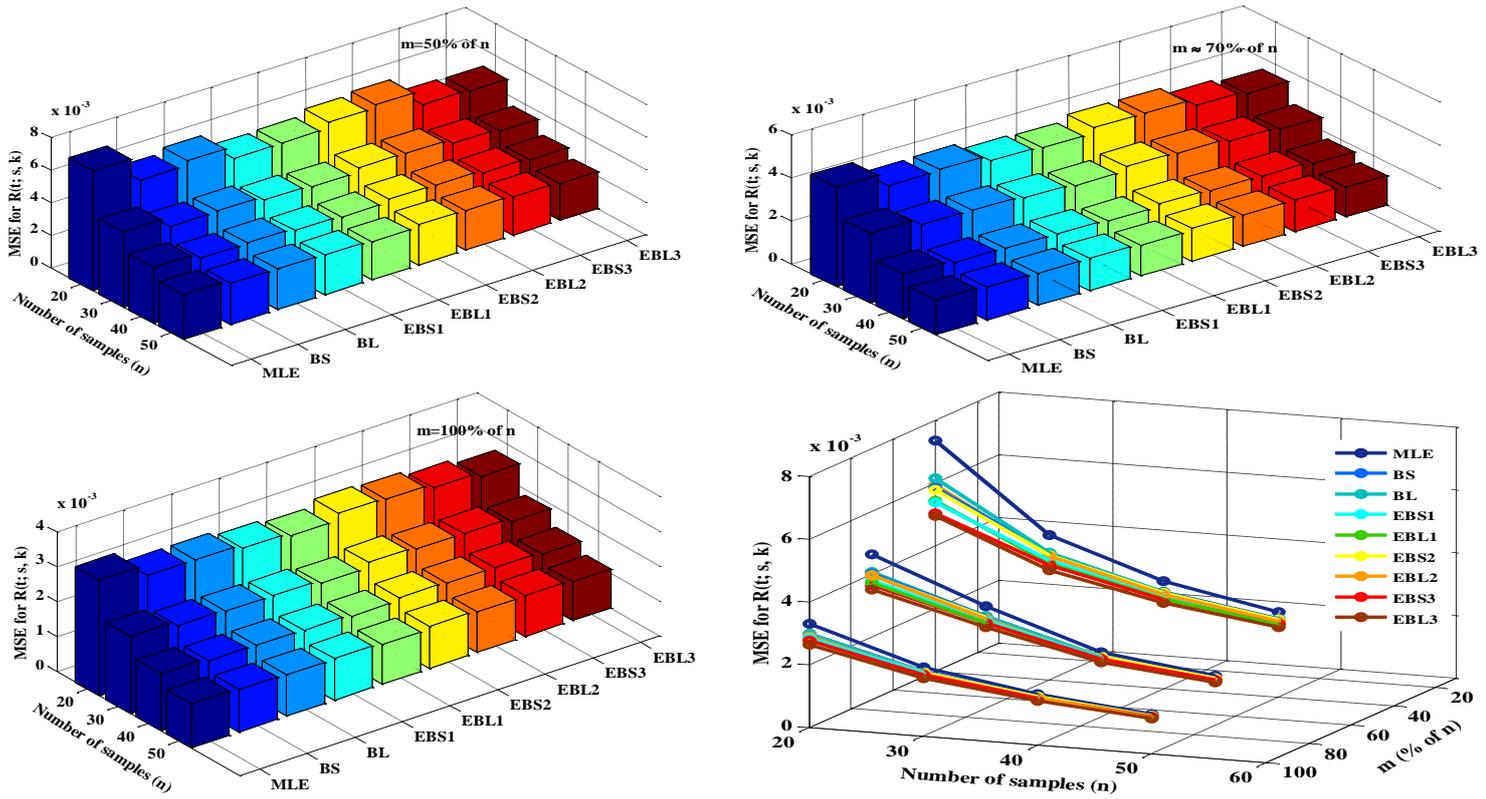

**Figure 2:** MSEs of estimations of R(t: s, k) based on (50, 70 and 100)% of samples.

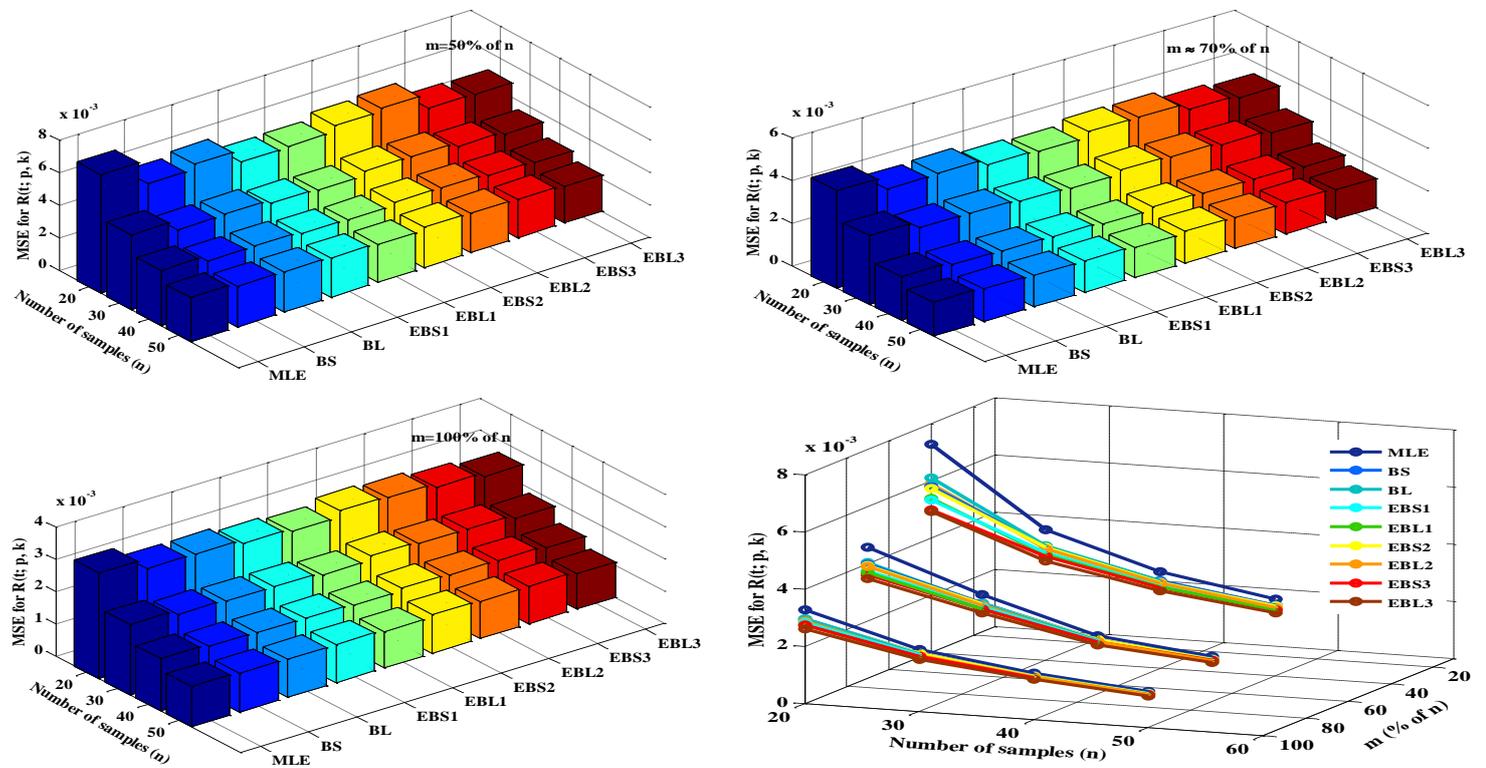

**Figure 3:** MSEs of estimations of R(t: p, k) based on (50, 70 and 100)% of samples.

**Table 7.** Estimates of $h(t)$ and their *MSE* based on WGED and Monte Carlo simulation ($q$=1, $t$=0.1, $h(t)$= 2.005066).

| n, m ,R | MLE | BS | BL | EBS1 | EBL1 | EBS2 | EBL2 | EBS3 | EBL3 |
|---|---|---|---|---|---|---|---|---|---|
| n=20, m=10; R=(4,4,2,0*7) | 2.216174 | 2.187792 | 2.076031 | 2.094185 | 1.987226 | 2.137266 | 2.025615 | 2.051104 | 1.948838 |
|  | 0.674473 | 0.558081 | 0.421966 | 0.488749 | 0.382106 | 0.543880 | 0.415408 | 0.439549 | 0.353220 |
| n=20, m=15; R=(2,2,1,0*12) | 2.168576 | 2.154944 | 2.082045 | 2.091854 | 2.021158 | 2.119557 | 2.046905 | 2.064151 | 1.99541 |
|  | 0.389166 | 0.349479 | 0.288322 | 0.31508 | 0.265851 | 0.338484 | 0.282022 | 0.293734 | 0.251416 |
| n=20, m=20; R=(0*20) | 2.117449 | 2.109788 | 2.057064 | 2.062917 | 2.011413 | 2.082834 | 2.030301 | 2.043001 | 1.992526 |
|  | 0.269908 | 0.250912 | 0.218479 | 0.232298 | 0.205944 | 0.244438 | 0.214759 | 0.221148 | 0.198008 |
| n=30, m=15; R=(5*3,0*12) | 2.144446 | 2.132216 | 2.061065 | 2.069812 | 2.000811 | 2.096824 | 2.025954 | 2.042800 | 1.975669 |
|  | 0.350029 | 0.316553 | 0.26392 | 0.286692 | 0.245287 | 0.306644 | 0.258773 | 0.268639 | 0.233414 |
| n=30, m=20; R=(4,4,2,0*17) | 2.112598 | 2.105213 | 2.052813 | 2.05845 | 2.007263 | 2.078239 | 2.026036 | 2.038660 | 1.988490 |
|  | 0.25701 | 0.239259 | 0.20865 | 0.221581 | 0.196938 | 0.233001 | 0.205151 | 0.211127 | 0.189584 |
| n=30, m=30; R=(0*30) | 2.078018 | 2.07444 | 2.040412 | 2.043372 | 2.00988 | 2.056162 | 2.022241 | 2.030582 | 1.997518 |
|  | 0.155865 | 0.149238 | 0.136218 | 0.141384 | 0.130784 | 0.146151 | 0.134378 | 0.136992 | 0.127539 |
| n=40, m=20; R=(5*4,0*16) | 2.116883 | 2.109333 | 2.056709 | 2.062475 | 2.011069 | 2.082352 | 2.029921 | 2.042598 | 1.992216 |
|  | 0.261896 | 0.243359 | 0.211665 | 0.225131 | 0.199472 | 0.236986 | 0.208039 | 0.214258 | 0.191777 |
| n=40, m=30; R=(4,4,2,0*27) | 2.079655 | 2.076004 | 2.041891 | 2.044911 | 2.011334 | 2.057734 | 2.023726 | 2.032088 | 1.998942 |
|  | 0.161421 | 0.154446 | 0.140894 | 0.146333 | 0.135224 | 0.151305 | 0.139001 | 0.141740 | 0.131800 |
| n=40, m=40; R=(0*40) | 2.0558 | 2.05365 | 2.028583 | 2.030459 | 2.00569 | 2.039862 | 2.014860 | 2.021056 | 1.996521 |
|  | 0.113279 | 0.109795 | 0.102726 | 0.105537 | 0.099759 | 0.108103 | 0.101731 | 0.103166 | 0.097972 |
| n=50, m=25; R=(5*5,0*20) | 2.093179 | 2.08823 | 2.046914 | 2.050867 | 2.010324 | 2.066425 | 2.025251 | 2.035309 | 1.995398 |
|  | 0.193771 | 0.183487 | 0.16431 | 0.172109 | 0.156557 | 0.179183 | 0.161796 | 0.165606 | 0.151839 |
| n=50, m=40; R=(4,4,2,0*37) | 2.057082 | 2.054916 | 2.029822 | 2.031711 | 2.006915 | 2.041124 | 2.016094 | 2.022297 | 1.997735 |
|  | 0.112958 | 0.10951 | 0.102421 | 0.105203 | 0.099406 | 0.107777 | 0.101386 | 0.102826 | 0.097613 |
| n=50, m=50; R=(0*50) | 2.045462 | 2.043982 | 2.024115 | 2.025459 | 2.005781 | 2.032902 | 2.013079 | 2.018015 | 1.998484 |
|  | 0.084991 | 0.082961 | 0.078642 | 0.080312 | 0.076791 | 0.081870 | 0.077995 | 0.078873 | 0.075702 |

From Table 7, the MSEs of all estimates decrease as *n* increases. The MSEs of estimates are ordered as follows:

$$\text{MSE}(\hat{h}_{EBs3}) < \text{MSE}(\hat{h}_{EBs1}) < \text{MSE}(\hat{h}_{EBs2}) < \text{MSE}(\hat{h}_{Bs}) < \text{MSE}(\hat{h}_{ML}),$$

$$\text{MSE}(\hat{h}_{EBL3}) < \text{MSE}(\hat{h}_{EBL1}) < \text{MSE}(\hat{h}_{EBL2}) < \text{MSE}(\hat{h}_{BL}) < \text{MSE}(\hat{h}_{ML}).$$

The E-Bayesian estimates have smaller MSEs than the Bayesian estimates. By increasing n, the E-Bayesian estimates have the smallest MSEs compared to their corresponding Bayesian and ML estimates and can be seen in Figure 4.

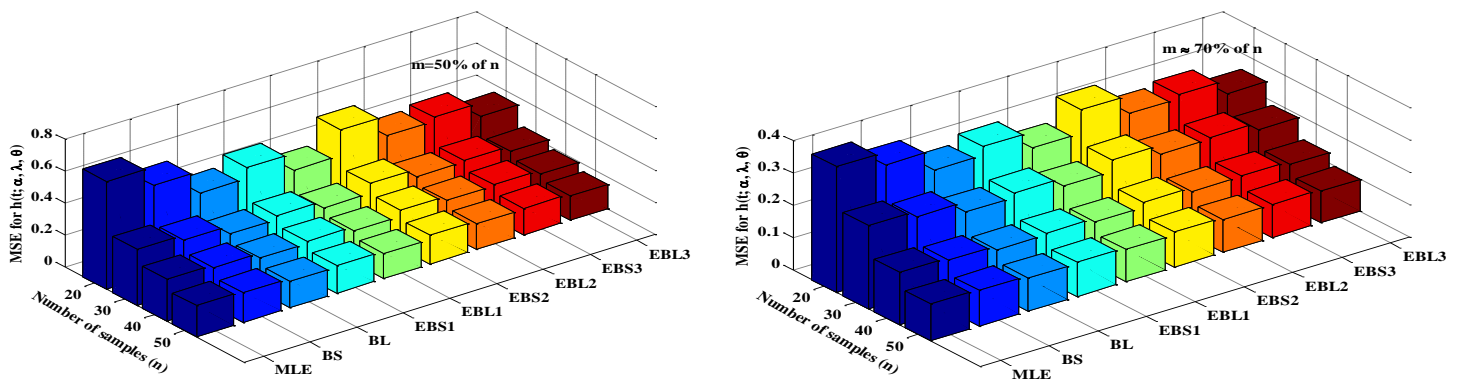

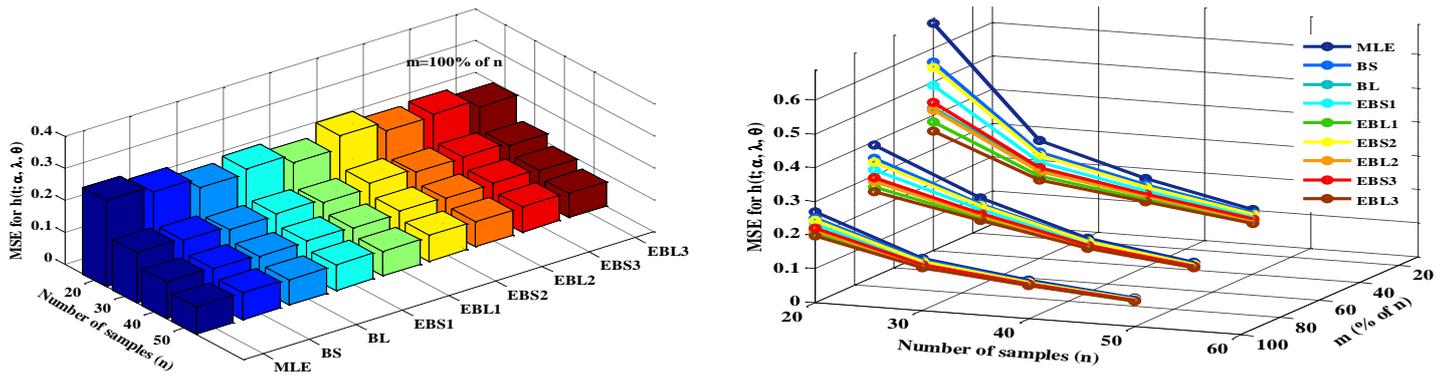

**Figure 4:** MSEs of estimations of h(t) based on (50, 70 and 100)% of samples.

## 7. CONCLUSIONS

In this paper, the MLE, Bayes, and E-Bayesian techniques are used for estimating the parameter, reliability, and hazard functions of the WGED based on progressive type-II censoring. We are used real data and the Monte Carlo simulation to compute and comparison of estimates. Also, we have found that:

- The E-Bayesian estimates of different priors are more nearly as the sample size increases. By different prior distributions of hyperparameters, the E-Bayesian estimations are robust and satisfy the corresponding theorems.
- The MSEs of estimates decreases as the sample size increases.
- The MSEs of the E-Bayesian estimates were less than the MSEs of their corresponding MLE and Bayes estimates.
- Generally, this paper shows that the E-Bayesian estimation is more efficient than the old methods.

***Declarations of interest***: none

***Funding Sources***: This research did not receive any specific grant from funding agencies in the public, commercial, or not-for-profit sectors.

**Appendices**

**Appendix A: Proof of Theorem 5.1.**

(1) From (4.3)-(4.5), we have:

$$\hat{\alpha}_{EBs2} - \hat{\alpha}_{EBs3} = 2(\hat{\alpha}_{EBs1} - \hat{\alpha}_{EBs3}),$$

and

$$\hat{\alpha}_{EBs1} - \hat{\alpha}_{EBs3} = \hat{\alpha}_{EBs2} - \hat{\alpha}_{EBs1} = \frac{1}{c}\left(m + \frac{u}{u+v}\right)\left[\frac{c+2S_m}{c}\ln\left(\frac{c+S_m}{S_m}\right) - 2\right]. \quad (A.1)$$

Now, according to Mac laurin series, we have:

$$\ln(1+x) = x - \frac{x^2}{2} + \frac{x^3}{3} - \frac{x^4}{4} + \cdots = \sum_{k=1}^{\infty}(-1)^{k-1}x^k/k, \quad |x| < 1. \quad (A.2)$$

Now, it suffices to replace $x$ by $c/S_m$ in (A.2). Also, note that when $0 < c < S_m$, then we have $0 < c/S_m < 1$, and:

$$\left[\frac{c+2S_m}{c}\ln\left(\frac{S_m+c}{S_m}\right) - 2\right]$$
$$= \frac{c+2S_m}{c}\left((c/S_m) - \frac{1}{2}(c/S_m)^2 + \frac{1}{3}(c/S_m)^3 - \frac{1}{4}(c/S_m)^4 + \frac{1}{5}(c/S_m)^5 - \cdots\right) - 2$$
$$= \left((c/S_m) - \frac{1}{2}(c/S_m)^2 + \frac{1}{3}(c/S_m)^3 - \frac{1}{4}(c/S_m)^4 + \frac{1}{5}(c/S_m)^5 - \cdots\right) - 2$$
$$+ \left(2 - (c/S_m) + \frac{2}{3}(c/S_m)^2 - \frac{2}{4}(c/S_m)^3 + \frac{2}{5}(c/S_m)^4 - \cdots\right)$$
$$= \left(c^2/6S_m^2 - c^3/6S_m^3\right) + \left(3c^4/6S_m^4 - 2c^5/15S_m^5\right) + \cdots$$
$$= \frac{c^2}{6S_m^2}(1 - c/S_m) + \frac{c^4}{60S_m^4}(9 - 8c/S_m) + \cdots. \quad (A.3)$$

Now according to (A.1) and (A.3), we have:

$$\hat{\alpha}_{EBs1} - \hat{\alpha}_{EBs3} = \hat{\alpha}_{EBs2} - \hat{\alpha}_{EBs1} > 0, \quad \text{or:} \quad \hat{\alpha}_{EBs3} < \hat{\alpha}_{EBs1} < \hat{\alpha}_{EBs2}.$$

(2) From (A.1) and (A.3), we get:

$$\lim_{S_m \to \infty}(\hat{\alpha}_{EBs1} - \hat{\alpha}_{EBs3}) = \lim_{S_m \to \infty}(\hat{\alpha}_{EBs2} - \hat{\alpha}_{EBs1}) = 0,$$

or: $\lim_{S_m \to \infty} \hat{\alpha}_{EBs1} = \lim_{S_m \to \infty} \hat{\alpha}_{EBs2} = \lim_{S_m \to \infty} \hat{\alpha}_{EBs3}$. ∎

**Appendix B: Proof of Theorem 5.2.**

(1) From (4.9)-(4.11), we have:

$$\hat{R}_{EBps3} - \hat{R}_{EBps1} = \hat{R}_{EBps1} - \hat{R}_{EBps2}$$

$$= \frac{2b-c}{c^2 B(u,v)} \int_0^1 \int_0^c \sum_{i=1}^{k}(-1)^{i-1}\binom{k}{i}\left(\frac{b+S_m}{b+S_m+i(\exp(\lambda t)-1)^\theta}\right)^{m+a} a^{u-1}(1-a)^{v-1}dbda \quad (B.1)$$

The equation (B.1) cannot be computed analytically in a simple closed form. Thus numerical methods by the corresponding software are used for computing it. The numerical results (Table 5 and Table 6) showed that this integral is positive. That is:

$$\hat{R}_{EBps2} < \hat{R}_{EBps1} < \hat{R}_{EBps3}.$$

Similarly, we have: $\hat{R}_{EBss2} < \hat{R}_{EBss1} < \hat{R}_{EBss3}.$

(2) From (B.1) we have

$$\lim_{S_m \to \infty}\left(\hat{R}_{EBps1} - \hat{R}_{EBps3}\right) = \lim_{S_m \to \infty}\left(\hat{R}_{EBps2} - \hat{R}_{EBps1}\right)$$

$$= \frac{c - 2b}{c^2 B(u,v)} \lim_{S_m \to \infty} \int_0^1 \int_0^c \sum_{i=1}^k (-1)^{i-1} \binom{k}{i}\left(\frac{b + S_m}{b + S_m + i(\exp(\lambda t) - 1)^\theta}\right)^{m+a} a^{u-1}(1-a)^{v-1} db\, da.$$

Now using the monotone convergence theorem, we have:

$$\lim_{S_m \to \infty} \hat{R}_{EBps1} = \lim_{S_m \to \infty} \hat{R}_{EBps2} = \lim_{S_m \to \infty} \hat{R}_{EBps3}.$$

Similarly, we have:

$$\lim_{S_m \to \infty} \hat{R}_{EBss1} = \lim_{S_m \to \infty} \hat{R}_{EBss2} = \lim_{S_m \to \infty} \hat{R}_{EBss3}. \blacksquare$$

**Appendix C: Proof of Theorem 5.3.**

(1) From (4.12)-(4.14) and (4.3)-(4.5), we get:

$$\hat{h}_{EBs1} - \hat{h}_{EBs3} = \hat{h}_{EBs2} - \hat{h}_{EBs1} = (1/c)\lambda\theta\exp(\lambda t)(\exp(\lambda t) - 1)^{\theta-1}\left(m + \frac{u}{u+v}\right)\left[\frac{c + 2S_m}{c}\ln\left(\frac{c + S_m}{S_m}\right) - 2\right]. \quad (C.1)$$

Now, according to (C.1), (A.1) and (A.3), we have:

$$\hat{h}_{EBs1} - \hat{h}_{EBs3} = \hat{h}_{EBs2} - \hat{h}_{EBs1} > 0, \quad \text{or:} \quad \hat{h}_{EBs3} < \hat{h}_{EBs1} < \hat{h}_{EBs2}.$$

(2) According to (C.1), (A.1) and (A.3), we have:

$$\lim_{S_m \to \infty}(\hat{h}_{EBs1} - \hat{h}_{EBs3}) = \lim_{S_m \to \infty}(\hat{h}_{EBs2} - \hat{h}_{EBs1})$$

$$= (1/c)\lambda\theta\exp(\lambda t)(\exp(\lambda t) - 1)^{\theta-1}\left(m + \frac{u}{u+v}\right)\lim_{S_m \to \infty}\left\{\frac{c^2}{6S_m^2}(1 - c/S_m) + \frac{c^4}{60S_m^4}(9 - 8c/S_m) + \cdots\right\} = 0.$$

Or, $\lim_{S_m \to \infty} \hat{h}_{EBs1} = \lim_{S_m \to \infty} \hat{h}_{EBs2} = \lim_{S_m \to \infty} \hat{h}_{EBs3}. \blacksquare$